# Observation of a push force on the end face of a nm fiber taper exerted by outgoing light

Weilong She*, Jianhui Yu*, Raohui Feng

State Key Laboratory of Optoelectronic Materials and Technologies, Sun Yat-Sen University, Guangzhou 510275, China

* **Corresponding authors:** w. s. shewl@mail.sysu. edu.cn, j. y. kensom@fish-finder.org

There are two different proposals for the momentum of light in a transparent dielectric of refractive index *n*: Minkowski's version $nE/c$ and Abrahm's version $E/(nc)$, where $E$ and $c$ are the energy and vacuum speed of light, respectively. Despite many tests and debates over nearly a century, momentum of light in a transparent dielectric remains controversial. In this Letter, we report a direct observation of the inward push force on the end face of a free nm fiber taper exerted by the outgoing light. Our results clearly support Abraham momentum. Our experiment also indicates an inward surface pressure on a dielectric exerted by the incident light, different from the commonly recognized pressure due to the specular reflection. Such an inward surface pressure by the incident light may be useful for precise design of the laser-induced inertially-confined fusion.

The momentum of light in dielectrics is one of Rudolf Peierls' classic surprises in theoretical physics [1,2]. In a transparent dielectric of refractive index *n*, Minkowski suggested $nE/c$ for the momentum of light [3], whereas Abraham suggested $E/(nc)$ for the same physical quantity [4], where $E$ and $c$ are the energy and vacuum speed of light respectively. There are a significant number of irradiative theories to clarify these two momenta [5-22], the number of experimental tests are, however, far less [23-26]. One of the experimental tests observing the radiation pressure at the water-air interface [23] is now reckoned to support Abraham momentum [2]. On the other hand, the measurement of systematic shift in the photon recoil frequency due to the index of refraction of the Bose-Einstein condensate suggested that Minkowski momentum is true [24]. Furthermore, the experiment of the light pressure on a mirror suspended in water appeared to support Minkowski momentum [25], but the careful calculation of total transferred momentum of light beam in the form of a short single-photon pulse fitted Abraham's value [11]. Similar case was the experiment of photon drag effect in semiconductors [26, 13, 16]. Despite the tests and debates over nearly a century, the controversy of momentum of light in dielectrics remains [2]. The main difficulty is experimental identification of light momentum in a transparent dielectric [1,2]. The complexity of momentum transfer of light in the dielectric obstructs the direct observation of light momentum, making experimental interpretations ambiguous. The way to observe light momentum in a lossless transparent dielectric by exploring the light pressure on the surfaces of a parallel-sided block of dielectric in air or in vacuum was shown unpromising [9]. And the beautiful idea proposed by Ashkin and Dziedzic on the experiment of radiation pressure at the single water-air interface [23] also encountered difficulty. Ashkin and Dziedzic's idea was very simple and perspicuous: when light passes through the interface, the resulting force will move the interface back if the Abraham momentum is correct, and the movement of the interface will be along the opposite direction if the Minkowski momentum is correct. In their experiment, Ashkin and Dziedzic observed an outward bulge on the interface, which appears to support Minkowski momentum. However, the bulge of water surface was found to be due to the lateral dipole forces caused by the intensity gradient of a Gaussian beam, which conceals the effect of light momentum [6, 13]. We have found a way to overcome this difficulty by replacing the water surface by a nm silica fiber taper (FT) and letting light travel in the FT then emerge into vacuum or air from the free end of the FT. In this Letter we report direct observation of an inward push force on the end face of the FT exerted by the outgoing light. Our experimental results clearly support Abraham momentum.

To illustrate our idea, we briefly review the radiation pressure on the surface of a lossless nonmagnetic transparent dielectric in vacuum or in air. If the refractive index of the dielectric $n > 1$ and the incident light beam is normal to the surface of the dielectric, the intensity reflectivity is given by $R \,[= (n-1)^2/(n+1)^2]$. By using Abraham momentum, the light will have a momentum change of magnitude $E/c \cdot (1-R)(1-1/n)$ due to transmission at the surface, where $E$ is the energy of the incident light; the momentum change due to the specular reflection is $2RE/c$ [or $2RE/(nc)$] when it arrives at the surface from outside (or inside) the dielectric. Therefore, when a light wave enters (or leaves) the dielectric the total decrease (or increase) of its momentum at the surface is $2E/c \cdot (n-1)/(n+1)$ [or $2E/(nc) \cdot (n-1)/(n+1)$], in agreement with Loudon's theory based on the Lorentz force [11]. By conservation of momentum, the surface of the dielectric will gain (or lose) a momentum $2E/c \cdot (n-1)/(n+1)$ [or $2E/(nc) \cdot (n-1)/(n+1)$] from (or to) the light and experience an *inward* radiation pressure for both cases of the light wave entering and leaving. Using Minkowski momentum in similar fashion, we can also deduce that, when a light wave enters (or leaves) the dielectric, the surface will have a momentum change of $2E/c \cdot (n-1)/(n+1)$ [or $2nE/(c) \cdot (n-1)/(n+1)$] pointing to vacuum or air, and experience an *outward* radiation pressure. Therefore, when a (suitably strong) light wave travels through a small silica FT and emerges into vacuum or air from the free end of the FT, we expect that the taper end face will experience a force and be compelled to move backward if Abraham momentum applies or forward if Minkowski momentum applies. We also expect that the FT's response will be expeditious due to its light weight and flexibility.

Figure 1 shows the experimental setup for observation of optical force, where 650 nm and 980 nm lights are from a weak CW semiconductor laser with a fixed output of 10 mW and a strong unpolarized CW fiber laser with a tunable output of 0-280 mW. They are coupled into the FT synchronously by a fiber coupler. The 980 nm light is used to drive the FT tail to move while the red 650 nm light is used to highlight the image of FT tail since it is too dim to observe when only with the scattering of 980 nm light. The power of the 650 nm light coupled into the FT is 0.5 mW and that of the 980 nm light is from 0 to 78 mW tunable. The coupled-in power of 980 nm light has been calibrated before the experiment. The FT is mounted in a hermetical flat-circinal-shape glass chamber with a diameter of 10.5 cm to avoid the influence of the flowing air. The space from the top (or bottom) of the glass chamber to the FT is about 1 cm. We erect the glass chamber and let the FT come down naturally like that shown in Fig. 2 (a). The movement of the FT tail is taken by a digital camera, Canon Power Shot A520, which is set at Movie mode with a rate of 10 frames/s and a resolution of $640 \times 480$ pixels. A convex lens is inserted between the glass chamber and the camera for taking the suitably large, clear image of the movement of the FT tail.

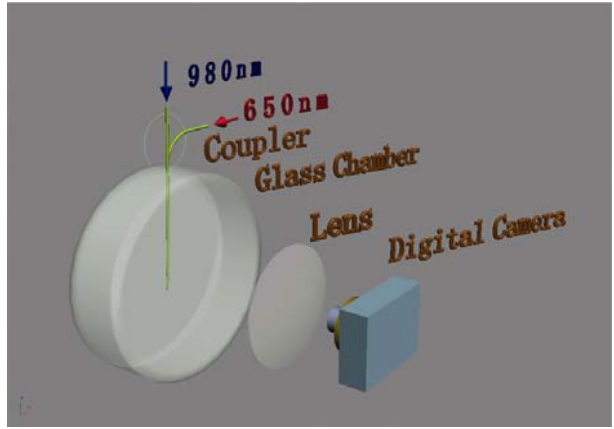

**Fig. 1 The experimental setup for observation of optical force. The light sources are a weak CW 650 nm semiconductor laser and a strong unpolarized CW 980 nm fiber laser. Their outputs are coupled into the fiber taper synchronously by a fiber coupler. The 980 nm light is used to drive the FT tail to move while the red 650 nm light is used to highlight the image of FT tail. The power of the 650 nm light coupled into the fiber taper is 0.5 mW fixed and that of the 980 nm light is from 0 to 78 mW tunable. The movement of the fiber taper tail driven by radiation pressure is taken by a digital camera (Canon Power Shot A520), which is set at Movie mode with a rate of 10 frames/s**.

The fiber tapers used are fabricated with single-mode fibers (SMF-28, from Corning Company), by a technique similar to that reported in [27]. Figure 2 elucidates the stationary imaging of a typical FT. Figure 2 (a) is the image of the FT taken by an optical system constructed by a quartz convex lens (4.2 cm focus length) and a digital camera (Canon G5). The top part with the largest dimension is the single-mode fiber taken off the outermost envelope, with a diameter of 125 μm. And the insert in Fig. 2 (a) is the repeated image of the FT tail but with the 4 mW and 650 nm light traveling in it, showing that the tail with scattering is about 6 mm. The diameter of the tail is measured by a microscope (Hisomet II DH2 series from Union Optical Co., Ltd) elsewhere. The arrows indicated as 1, 2 and 3 correspond to diameters equal to 4.7, 2.5 and 1.2 μm, respectively. Figure 2 (b) is the micrograph of the thin tip of FT with 500 nm diameter. The insert in Fig. 2 (b) is the micrograph of FT end face with a very weak outgoing red light, taken by the same microscope, showing that the end face is a circular one. It should be noticed that Fig. 2 is taken after the experiment described below. The fact that the tail of the FT in Fig. 2 looks slightly different from that in Fig. 3 is due to the flexibility of FT.

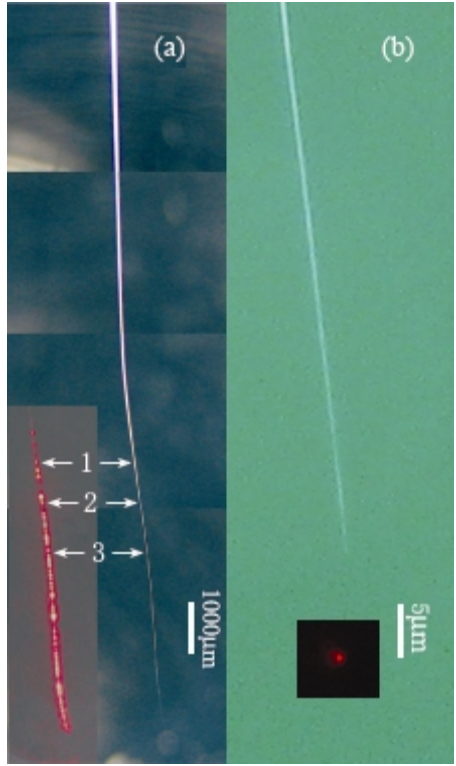

**Fig. 2 The stationary imaging of FT used. (a) The image of the FT taken by an optical system constructed by a quartz convex lens (4.2 cm focus length) and a digital camera (Canon G5): the top part with the largest dimension is the single-mode fiber taken off the outermost envelope; the insert in (a) is the repeated image of the FT tail but with a 4 mW and 650 nm light traveling in it. The diameters of FT indicated by arrows 1, 2 and 3 are 4.7, 2.5 and 1.2 μm, respectively, measured by a microscope (Hisomet II DH2 series from Union Optical Co., Ltd). (b) The micrograph of the thin tip of FT with 500 nm diameter, where the insert in (b) is the micrograph of FT end face with a very weak outgoing red light, taken by the same microscope.**

During experiment with FT shown in Fig.2, we first turn on the digital camera on at zoom-in mode, then tune the power of the 980 nm laser slowly and observe the pose of the FT end carefully. We find that when the power coupled in the FT is 4-5 mW the taper end shows a slight action, but when the coupled-in power is about 50 mW the FT tail begins to move, making an oscillation. Figure 3 displays video frames of the oscillating FT tail as the result of the optical force, where 3 (a), (b), and (c) are the images of the FT tail corresponding respectively to 0, 4 and 6 mW coupled-in the 980 nm light. Figure 3 (d)-(x) are those selected from another movie, corresponding to the times of 0.3, 3.0, 3.9, 4.9, 5.6, 6.1, 6.7, 7.0, 7.3, 7.9, 8.3, 8.6, 8.7, 8.9, 9.3, 9.5, 9.7, 10.2, 10.6, 11.1 and 11.3 s, respectively; the origin of time is the moment when 50 mW of the 980 nm light is just coupled in. The poses of the FT shown in Figs. 3 (a)-(x) can also be observed when the coupled-in power of the 980 nm light is tuned from 0 to 50 mW continuously [see the video in AIP's Electronic Physics Auxiliary Publication Service]. Comparing Figs. 3 (a), (b) and (c), we can find that the poses of the thin tip of the FT, indicated by three arrows, are different from each other; the cases of (a) and (c) are especially so. From Figs. 3 (a) to (x), we can also find that the FT tail is driven to move by a push force on the end of the FT. Figures 3 (a) to (f) show that the tip of the FT with a length approximately 0.7 mm is pushed to right, up and away from reader; the brightest part of the FT near the tip moves left, towards reader. Figures 3 (g) to (j) show that the tip turns left and the section of the FT lower than the arrow in (h) is bent left, then comes back, due to the push force. Figures 3 (k) to (p) show that the FT tail swings towards right, driven by the push force with (p) corresponding to the limiting position, where the tip points right and downward. Figures 3 (q) to (w) show that the tip is driven to left again. Figure 3 (x) shows repetition of (l). We find that (l) to (x) is the period of motion and the subsequent motion just repeats earlier steps (see the video in AIP's Electronic Physics Auxiliary Publication Service). When the 980 nm laser is turned off, the moving FT comes back to the (a) state in about 2.5 s; if the laser is turned on again, the FT starts to move again. The same phenomenon is also observed with other FTs in air and a FT in a vacuum of $4\times10^{-5} Torr$ at $15°C$. The latter shows that the phenomenon observed is independent of air.

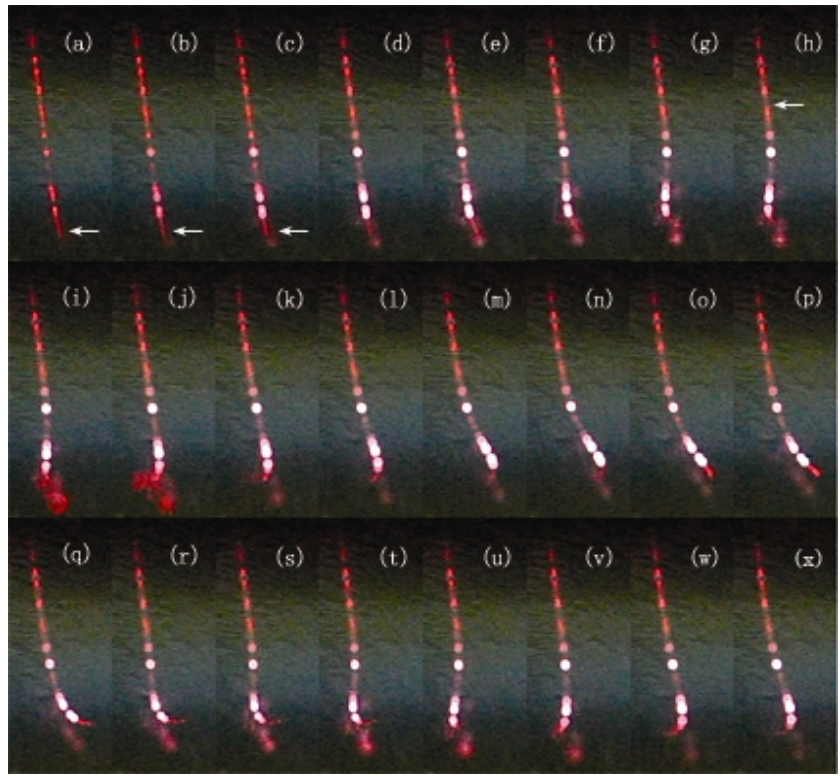

**Fig. 3 The displaying of video frames of the moving FT tail brought into oscillation by the push force of the 980 nm outgoing light. (a), (b), and (c) correspond respectively to 0, 4 and 6 mW coupled-in power of the light. And (d)-(x) correspond to 50 mW coupled-in power, at the times of 0.3, 3.0, 3.9, 4.9, 5.7, 6.1, 6.7, 7.0, 7.3, 7.9, 8.2, 8.6, 8.7, 8.9, 9.3, 9.5, 9.7, 10.2, 10.6, 11.1, and 11.3 s, respectively; the original point of time is the moment when 50 mW and 980 nm light is coupled in.**

The experiment described above shows clearly that the movement of the FT tail is due to the push force on the end face of the FT exerted by the outgoing light. The nature of movement cannot be explained by Minkowski momentum since Minkowski momentum predicts a pull force, pulling the tip straight rather than pushing it moving to right, up and away from reader as shown in Figs. 3 (a)-(f) and then oscillating as shown in Figs. 3 (g)-(x). Instead, we use Abraham momentum to analyze these movements. We focus on Figs. 3 (a), (b), and (c). We can see that when the power of the 980 nm light coupled in the FT is lower than 6 mW, the FT mostly is stationary except for its tip. So under this condition we can consider approximately that only the tip is moveable. When the push force is equal to the gravity on the tip, the tip is in a state of weightlessness; the equilibrium between gravity and elastic force of the tip is broken. And as the push force further increases, the tip begins to move. The push force can be deduced based on the momentum change of the light at the surface of the dielectric as discussed earlier, thus giving the value $f_A = 2P/(nc) \cdot (n-1)/(n+1)$, where P is the power of the outgoing light. On the other hand, the gravity on the tip is $f_g = \pi r^2 L \rho g$, where $r$, $L$ and $\rho$ are the radius, length and density of the tip, respectively; $g$ is the acceleration of gravity. When the push force is equal to the gravity, i.e., $f_A = f_g$, the light power is $P = \pi r^2 L \rho g n c (n+1)/[2(n-1)]$. For our FT tip, $r = 250 nm$, $L = 0.7 mm$, $\rho = 2.2 g/cm^3$ [28] and $n = 1.451$ according to Sellmeier formula [29]. We thus obtain $P = 3.51 mW$. Furthermore, by considering the dependence of the group velocity $v_g$ of the light at fundamental mode on the diameter of the silica FT tip [29], the push force should be amended according to the formula $f_A = P/c \cdot [(1-R)(1-1/n_g) - 2R/n_g]$, where $n_g = c/v_g$. The group velocity is calculated as $v_g = 0.704c$ at 980 nm, according to [29]. We thus have a slightly larger value for the light power $P = 3.73 mW$. From Figs. 3 (a), (b) and (c), we observed that when the coupled-in power is 4.5-6.5 mW (4-6 mW of the 980 nm light plus 0.5 mW of the 650 nm light), the tip of the FT begins to move. We see that our calculation is consistent with the experiment.

In conclusion, our experiment and analysis suggest that Abraham momentum is correct. Furthermore, our experiment also suggests a potentially important application of the momentum of light. As discussed above, when light is incident on the surface of a transparent dielectric from vacuum, it exerts an inward pressure on the surface of the dielectric. This pressure is different to the commonly recognized one due to the specular reflection. Therefore, when a series of incoherent laser beams with almost constant diameters is simultaneously incident on the surface of a transparent dielectric ball in vacuum along the radial direction, the dielectric ball will shrink due to the effect of Abraham momentum. This effect may be useful for the precise design of the laser-induced inertially-confined fusion.

The authors wish to thank Prof. Limin Tong for imparting the technique of making nm fiber taper, and Dr. Yang Xian for language assistance.